\documentclass[aps, pra, showpacs, preprint]{revtex4-2}
\usepackage{amssymb,amsfonts,amsmath,mathtext,longtable,array}
\usepackage{xcolor}

\usepackage{graphicx}
\usepackage{dcolumn}
\usepackage{bm}
\usepackage{color, xcolor}
\usepackage{natbib}
\usepackage{appendix}
\usepackage{multirow}

\begin{document}

\title{Noise avalanche and its quantum quenching in bosonic chains with random off-diagonal disorder}

\author{Vladislav~Stefanov}
\affiliation{Institute of Applied Physics, University of Bern, Sidlerstrasse 5, CH-3012 Bern, Switzerland}

\author{Andre Stefanov}
\affiliation{Institute of Applied Physics, University of Bern, Sidlerstrasse 5, CH-3012 Bern, Switzerland}

\author{Lea Sirota}
\affiliation{School of Mechanical Engineering, Tel Aviv University, Tel Aviv 69978, Israel}
\author{Gregory Slepyan}
\affiliation{School of Electrical Engineering, Tel Aviv University, Tel Aviv 69978, Israel}
\author{Dmitri Mogilevtsev}
\affiliation{School of Electrical Engineering, Tel Aviv University, Tel Aviv 69978, Israel}
\affiliation{B.I.Stepanov Institute of Physics, NAS of Belarus, Nezavisimosti ave. 68, 220072 Minsk, Belarus}

\begin{abstract}
Here we discuss a phenomenon of  sharp increase in the photon number noise at initial stages of propagation in tight-binding bosonic chains with off-diagonal disorder. Such a "noise avalanche" occurs under classical coherent excitation of waveguides and leads to high super-thermal photon bunching. Additional classical excitation slows but cannot quench this noise avalanche.  However, an additional single-photon excitation stops the avalanche.
\end{abstract}

\maketitle

\section{Introduction}

From early times of studying coupling between closely situated waveguides, researches were actively discussing an influence of coupling randomness on the field propagation through waveguides \cite{1127763,6768711,6774108,1976OptQE...8..503E}. Among many important results obtained in these works, it was established that randomness in coupling between waveguides can eventually lead to asymptotically equal field energy distribution between the waveguides irrespective of initial states. This phenomenon later was shown to be very common for systems with noisy coupling (i.e., so called "off-diagonal disorder") generically described by nonlocal Hermitian Lindblad operators in the weak Markovian noise limit, so collective dephasing induced by such randomness was even termed as a "grinder" in the field of multy-particle localization  \cite{PhysRevLett.118.070402}. 

Recently it was discovered  that dynamics of such "grinders" can be far more interesting than just establishing homogeneous field energy distribution over the waveguide system.  Namely, "grinders" can induce rather counter-intuitive behavior of field correlations even for classical initial states. In particular, a coherent initial state excited in one of the waveguides of the linear next-neighbor coupled chain of bosonic single-mode waveguides can lead to so called "photonic thermalization gap": a photon-number distribution in a particular waveguide could jump to be a super-thermal  one and during dynamics tend to the thermal one \cite{2015NatPh..11..930K}. So, there is a region between the thermal and coherent statistics that some modes of the structure are not reaching. This interesting phenomenon gave rise to a number of interesting developments addressing an interplay between topological features of the structure and manifestation of the "photonic thermalization gap", losses, etc. \cite{PhysRevLett.122.013903,10.1063/1.4992018,Wang_2019}. It was also shown that coherence in "grinders" can flow like heat even when the energy is already "grinded", i.e., homogeneously  spread over the modes  \cite{PhysRevA.94.012116}. Quantumness reveal even more surprising properties of "grinders". For instance,  non-classical initial states can be driven by off-diagonal disorder toward entangled stationary states \cite{PhysRevA.94.012116}.

Here we report another counter-intuitive phenomenon  that can arise in a coupled bosonic systems with random off-diagonal disorder: a noise avalanche. The essence of this effects is exponential increasing of the photon bunching at the initial stage of dynamics with distance from the initially coherently excited waveguide. We discuss this effect using both the density matrix approach and more conventional  effective Hamiltonian approach for modal amplitudes. We show how the noise avalanche effect arises in systems with Markovian dynamics produced by weak off-diagonal random disorder, and confirm results with numerical modeling performed by the effective Hamiltonian approach. We also provide analytical estimates showing how the exponential scaling arises. 

We apply the master equation approach for situation when phenomena akin to the "photonic thermalization gap" are taking place. We show that their manifestations strongly depend on the nature of the off-diagonal disorder and possible correlations between real and imaginary  components of the noised coupling constants. In particular, we show that the "gap" in bunching might be present but not reaching the thermal value. Also, a manifestation of the noise avalanche can be affected by the nature of the off-diagonal disorder. For instance,  real noise of coupling constants leads to much higher bunching than complex noise with independently fluctuating real and imaginary parts of the coupling constants.  

We also discuss how the noise avalanche is affected by non-local initial excitation. Additional coherent state excitation at some waveguide makes  the avalanche start from this waveguide. But the avalanche is not quenched.  Whereas an additional single-photon excitation completely quenches the avalanche turning off the exponential bunching increase. Second-order correlation functions of the modes beyond the  one initially excited by just a single-photon start from zero value.  We demonstrate single-photon quenching of the noise avalanche with both the effective Hamiltonian approach and the master equation approach. 

The outline of this paper is as follows. In the second Section the model of the  coupled single-mode waveguides is described. Here both the master equation approach and  the effective Hamiltonian approach are discussed. In the third Section a simple two-waveguide structure is discussed in framed of both approaches. Here is is demonstrated how the "photonic thermalization gap" manifestation  depends on the nature of the off-diagonal disorder noise. In particular, it is shown that real fluctuating interaction constant leads to larger bunching and the complex of-diagonal noise with independently fluctuating imaginary and real parts. In the third Section multi-waveguide structures are considered and noise avalanche is demonstrated with both approaches. Also, some analytic estimates are provided. The fourth Section is devoted to quenching the noise avalanche with single-photons. The fifth Section is about possible practical uses and realizations of the noise avalanche.

\section{Model}

Here we consider a chain of bosonic modes of the same frequency with next-neighbour coupling. Such a system is schematically depicted in Fig.~\ref{fig0-4}(a). It is useful to note that this chain can be considered as a generalization of the much discussed Su–Schrieffer–Heeger (SSH) model \cite{PhysRevLett.42.1698} and is popular workhorse for discussing a plephora of transport effects and topological features (see, for instance, the book \cite{2016LNP...919.....A} and Refs. \cite{doi:10.1126/science.aar7709,doi:10.1080/00018732.2021.1876991,PhysRevLett.100.170506,eichel,bigger,PhysRevLett.102.065703,PhysRevX.5.021025,PhysRevB.107.054308,Peshko:23}  ). This chain model can be described by the following Hamiltonian 
\begin{eqnarray}
H=H_0+V(t),
\label{hamg}
\end{eqnarray}
where the time-independent non-random part $H_0$ describes the conventional unitary SSH chain of $N+1$ resonant coupled bosonic modes (we set $\hbar\equiv 1$ in the following discussion):
\begin{eqnarray}
H_0=\sum\limits_{j=1}^N (v_ja^{\dagger}_ja_{j+1}+v^*_ja^{\dagger}_{j+1}a_j),
\label{ham0}
\end{eqnarray}
with $a^{\dagger}_j$ and $a_j$ being bosonic creation and annihilation operators for the $j$-th mode of the chain; $v_j$ are the interaction constants. The off-diagonal random disorder affecting the chain  we take as described by the following random addition to the Hamiltonian (\ref{ham0}): 
\begin{eqnarray}
V(t)=\sum\limits_{j=1}^N (\mu_j(t)a^{\dagger}_ja_{j+1}+\mu^*_j(t)a^{\dagger}_{j+1}a_j),
\label{hamv}
\end{eqnarray}
where $\mu_j(t)$ are random zero-mean classical variables, 
\[\langle \mu_j(t)\rangle_c=0,\] 
where  $\langle\ldots \rangle_c$ denotes classical averaging over the noise realizations. We assume that noises between each pair of modes are independent, 
\[\langle \mu_j(t)\mu_k(\tau)\rangle_c=\langle \mu_j(t)\mu^*_k(\tau)\rangle_c=0, \quad \forall j\neq k.\]

\subsection{Effective Hamiltonian approach}
 
This approach is commonly used in modern photonics when one just assumes the modal state to be coherent ones and derives equations for modal amplitudes. In our case off-diagonal noise destroys coherence.  However, for each particular noise realizations the dynamics remains unitary. For a pure initial state $|\Psi(0)\rangle$ modal dynamics can be described by the following Schroedinger equation 
\begin{equation}
\frac{d}{dt}|\Psi(t)\rangle=-i(H_0+V_k(t))|\Psi(t)\rangle,
\label{schrod1}    
\end{equation}
where $V_k(t)$ is a $k$-th realization of the random off-diagonal disorder terms described by Eq.(\ref{hamv}). 

For classical initial states it is rather simple to deal with Eq.(\ref{schrod1}).
Denoting $\mu_{j}^{(k)}(t)$ the $k$-th realization of the process $\mu_j(t)$, one gets from Eq.(\ref{schrod1}) the following system of equations for the $k$-th realization of the modal amplitudes for initial coherent states of each mode
\begin{equation} \label{efs1}
\frac{d}{dt}\alpha_{j}^{(k)}(t)=-i(v_j+\mu_{j}^{(k)}(t))\alpha_{j+1}^{(k)}(t)-i(v_{j-1}^*+\mu_{j-1}^{*(k)}(t))\alpha_{j-1}^{(k)}(t).
\end{equation}
Such a system was considered in the original work on the "photonic thermalization gap" and associated with chiral symmetry   \cite{2015NatPh..11..930K}. Calculating sets of amplitudes for different realizations, one can estimate  correlation functions.

Of course, ideally, a similar procedure can be performed with the original equation (\ref{schrod1}) for arbitrary initial quantum states building a set of "quantum trajectory" wave functions for finding averages (\cite{1993LNPM...18.....C,Molmer:93}). However, with increasing of the chain length and the number of basis vectors necessary to describe states this quantum trajectory approach quickly becomes  unwieldy.  For instance, for $K$ initial photons in the chain of $N$ waveguides one needs  $N^K$ orthogonal components in a "quantum trajectory" wave function.

In some cases,  there are ways to circumvent this difficulty. For example, for initial states that can be represented as a mixture of just few coherent state projectors with positive and negative weights (for example, when one has just a single-photon initial state in one mode and coherent states of other modes), the system (\ref{efs1}) can also be successfully applied for finding correlations functions of an arbitrary order.  The way of doing that is described in the recent works \cite{PhysRevA.105.052206,https://doi.org/10.1002/qute.202300060}. Also, for just few initially excited modes in arbitrary states and for normally correlation functions of low orders one can quite efficiently  use the effective Hamiltonian approach by solving the Heisenberg equations for the operators $a_j$ for each noise realization (this is discussed in more details in the Section IV).

\subsection{Master equation approach}
In difference with the effective Hamiltonian approach, a feasible and useful master equation can be obtained only by some rather restrictive limitations on the noises $\mu_j$. However, as it will be seen below, the master equation allows for finding effects that could be really hard to capture with the effective Hamiltonian approach. Also, the master equation allows for analytic estimation and provides useful guidelines for more detailed analysis with help of the effective Hamiltonian approach. 

Let us derive in a standard way the Lindblad master equation for the problem (\ref{hamg},\ref{ham0}) taking the noise to be delta-correlated in time \cite{breuer2002theory}:
\begin{eqnarray}
\nonumber
\langle \mu_j^*(t)\mu_j(\tau)\rangle_c=\gamma_j\delta(t-\tau), \\
\label{avs}
\langle \mu_j(t)\mu_j(\tau)\rangle_c=\kappa_j\delta(t-\tau). 
\end{eqnarray}
 where $\delta(t-\tau)$ is the Dirac delta-function; $\kappa_j$ and $\gamma_j\geq 0$ are dephasing rates describing noise-induced coupling between $j$-th and $j+1$-th modes.  

The Hamiltonians (\ref{hamg},\ref{hamv}) with correlations (\ref{avs}) lead to the following master equation for the density matrix of the mode array
\begin{multline} \label{mast1}
   \frac{d}{dt}\rho=-i[H_0,\rho]-\sum\limits_{j=1}^N({D}_j\rho+\rho{D}_j)
+2\sum\limits_{j=1}^N (\kappa_jL_j\rho L_j+\kappa_j^*L_j^{\dagger}\rho L_j^{\dagger} +
\gamma_j(L_j\rho L_j^{\dagger}+L_j^{\dagger}\rho L_j)),
\end{multline}
where the Lindblad operators are $L_j=a^{\dagger}_ja_{j+1}$ and  
\begin{equation}
D_j=\kappa_jL_j^2+\kappa_j^*(L_j^{\dagger})^2 +
\gamma_j(L_j L_j^{\dagger}+L_j^{\dagger} L_j).
\label{lind1}
\end{equation}
Eqs.(\ref{mast1},\ref{lind1}) are well illustrating an influence of the nature of noise  that produces off-diagonal disorder. Indeed, for instance, for real disorder  assumed in Ref. \cite{2015NatPh..11..930K} ( $\mathrm{Im}\mu_j(t)=0$, $\forall j$), one necessarily has $\kappa_j=\gamma_j\geq 0$, and the master equation (\ref{mast1}) reduces to the following one
\begin{equation}
\frac{d}{dt}\rho=-i[H_0,\rho]+\sum\limits_{j=1}^N{\gamma_j}\mathcal{D}(L_j+L_j^{\dagger})\rho,
    \label{mast2}
\end{equation}
where the dissipator is $\mathcal{D}(X)\rho=2X\rho X^{\dagger}-X^{\dagger}X\rho-\rho X^{\dagger}X$.

For the case of having white circular noise (i.e., $\kappa_j=0$, $\gamma_j\geq 0$, $\forall j$)  the master equation (\ref{mast1}) reduces to 
\begin{equation}
\frac{d}{dt}\rho=-i[H_0,\rho]+\sum\limits_{j=1}^N{\gamma_j}\left(\mathcal{D}(L_j)\rho+\mathcal{D}(L_j^{\dagger})\rho\right).
    \label{mast3}
\end{equation}
As we will see later, different master equations (\ref{mast2}) and (\ref{mast3}) not unexpectedly lead to rather different dynamics of correlations. 

The master equation approach described here is able to provide one with analytical solutions and sets of rather simple equations for the low-order correlation functions. Also, it is able to describe evolution of correlations for non-classical initial states of the chain.

\begin{figure*}[ht]
    \begin{minipage}[b]{.3\textwidth}
        \centering
        (a)\\
        \includegraphics[width=\linewidth]{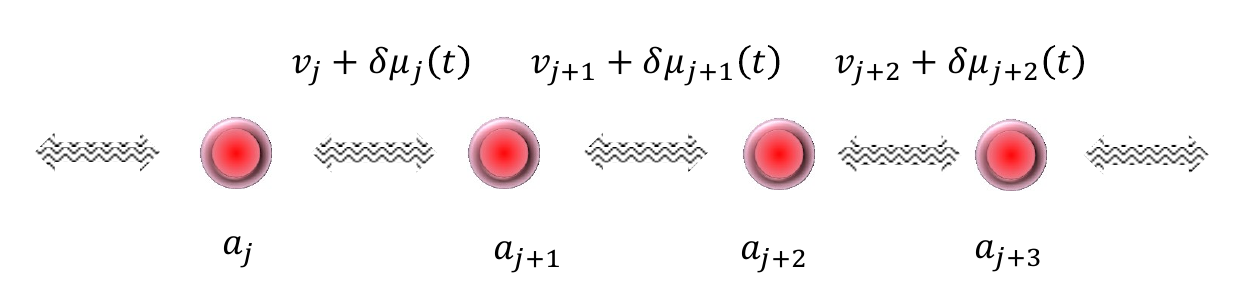}\\       
        \vspace{1cm}
        (b)\\
        \includegraphics[width=\linewidth]{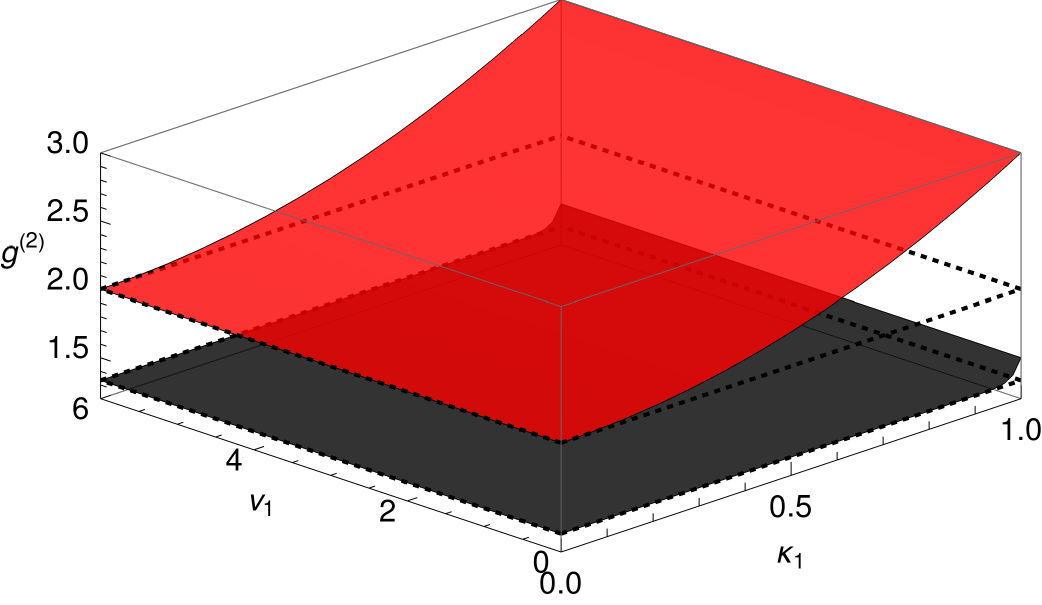}
    \end{minipage}
     \hfill
     \begin{minipage}[b]{.3\textwidth}
       \centering
        (c)\\
         \includegraphics[width=\linewidth]{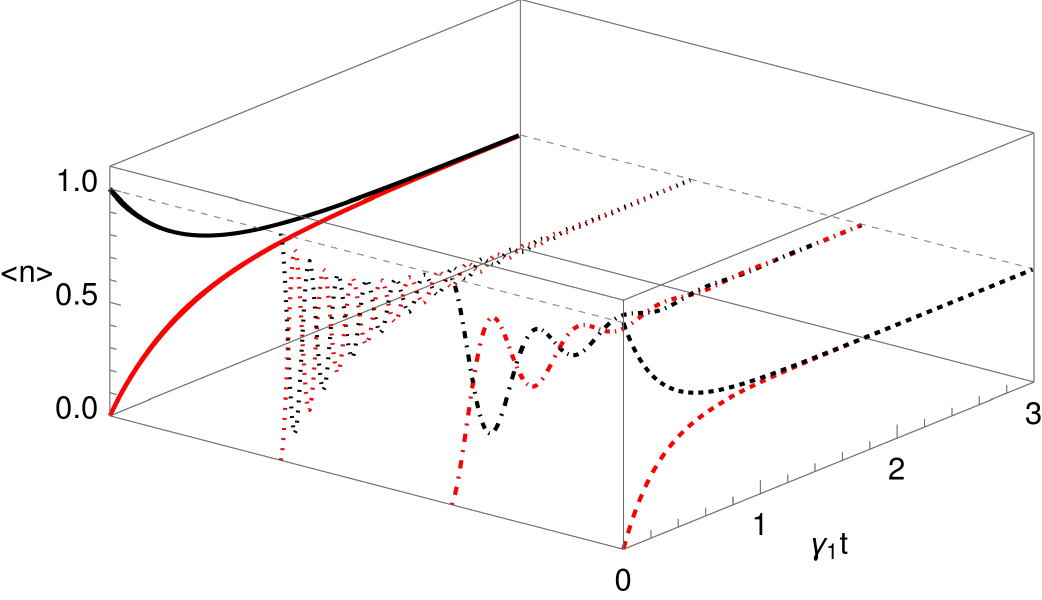}\\
        (d)\\
        \includegraphics[width=\linewidth]{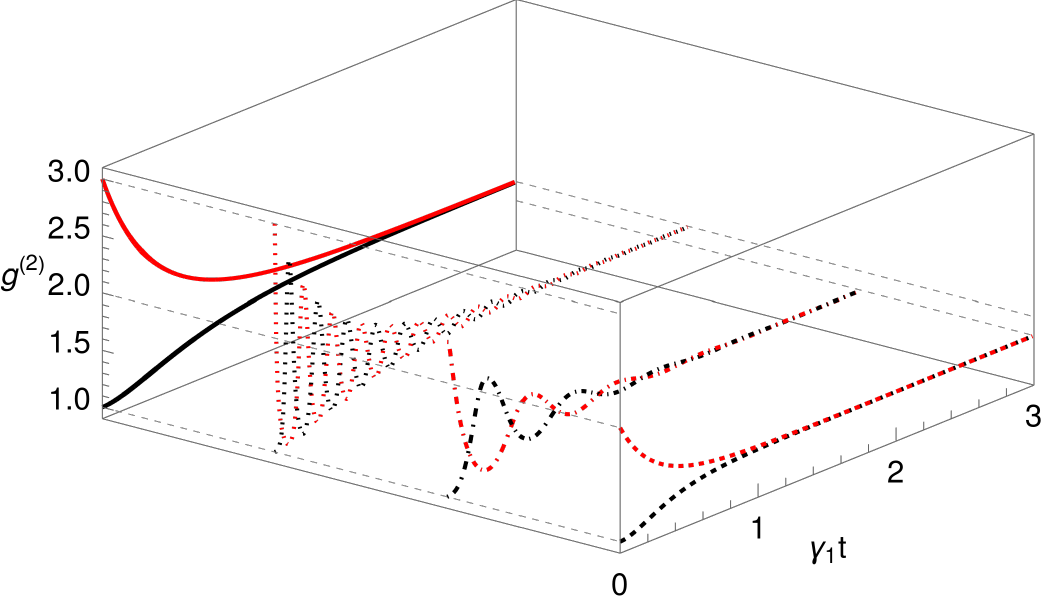}        
    \end{minipage}
    \hfill
    \begin{minipage}[b]{.3\textwidth}
        \centering
        (e)\\
        \includegraphics[width=\linewidth]{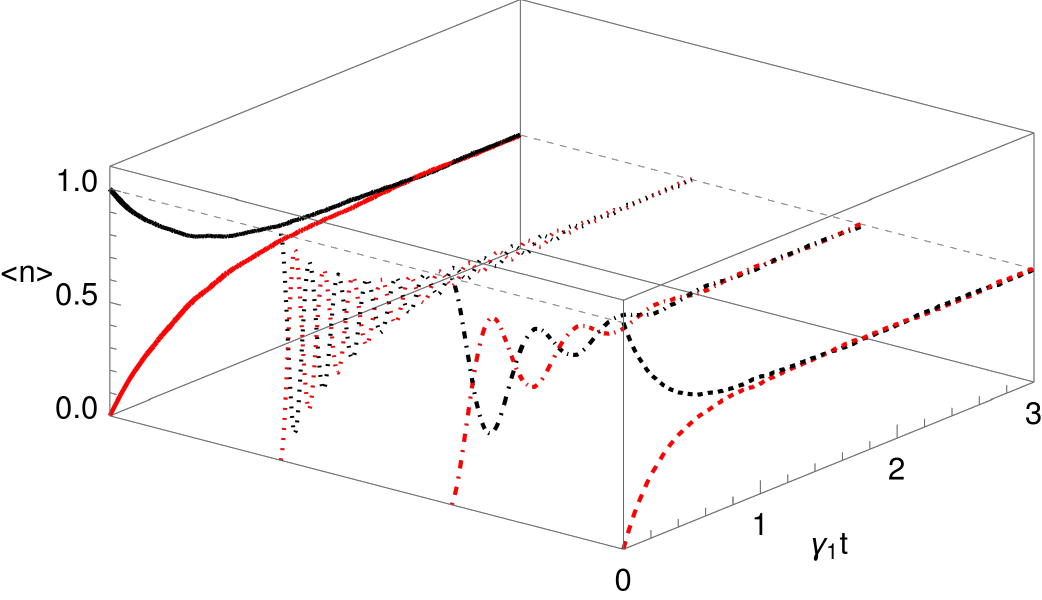}        
        (f)\\
        \includegraphics[width=\linewidth]{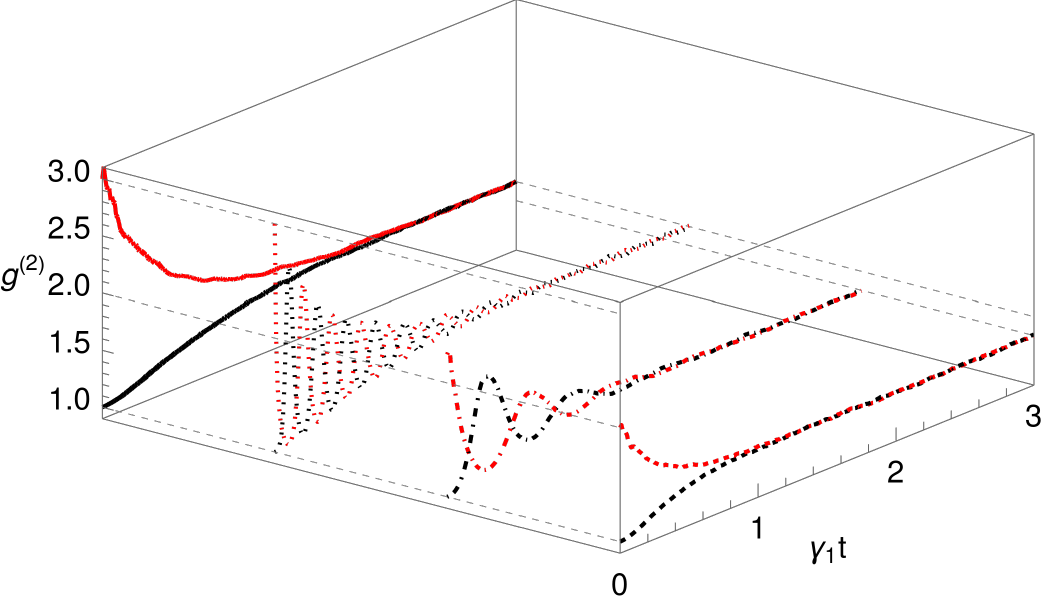}
    \end{minipage}    
\caption{(a) The scheme of the bosonic coupled chain with modes $a_j$ and corresponding couplings between them $v_j+\delta \mu_j(t)$. (b) Upper sheet shows $g^{(2)}_2(t)$ for $\gamma_1t=10^{-6}$ and different values of real $v_1$, $\kappa_1$. Lower sheet shows $g^{(2)}_2(t)$ for $\gamma_1 t=10$ and different values of real $v_1$, $\kappa_1$. The simulation for the panel (b) is obtained by the master equation approach.  (c) and (e) Average number of photons and (d) and (f) the normalized second-order correlation function in the mode $a_1$ (lower red curves) and in the mode $a_2$ (upper black curves) are drawn via the master equation (c, d) and  the effective Hamiltonian approaches (e, f).
In all panels (c,d,e,f) solid curves correspond to $v_1=0$, $\kappa_1=1$; dash-dotted curves correspond to $v_1=15\gamma_1$, $\kappa_1=1$; dashed curves correspond to $v_1=5\gamma_1$, $\kappa_1=3/4$;  dotted curves correspond to $v_1=0$, $\kappa_1=0$.
 The number of MC realization is $K=5*10^3$; 
    the number of time sub-intervals is $M=300$. For all the panels (b-f) the mode 1 is initially excited in the coherent state with the unit amplitude; the initial state of the mode 2 is vacuum.
 }
\label{fig0-4}
\end{figure*}

\section{Two-mode case}

To demonstrate functioning of both approaches, manifestation of  "grinding", "photonic thermal gap" and influence of noise character, let us consider first the simplest system of just two resonant modes with off-diagonal random disorder obtained for $N=2$ in Eqs.(\ref{ham0},\ref{hamv}). Here we discuss dynamics of average amplitudes, numbers of photons and second-order correlation functions. 

\subsection{Master equation approach for two modes}

For the average modal amplitudes one gets from Eqs.(\ref{ham0},\ref{hamv}) the following system of equations for average modal amplitudes 
\begin{eqnarray}
\frac{d}{dt}\vec{\phi}^{(1)}=-i\begin{pmatrix} 
-i\gamma_1 & v_1 \\
v_1^* & -i\gamma_1
\end{pmatrix}\vec{\phi}^{(1)},
\label{2am}
\end{eqnarray}
where the vector of amplitudes is $\left[\vec{\phi}^{(1)}\right]^T=[\langle a_1\rangle,\langle a_2\rangle]$. 
The system (\ref{2am}) describes just a unitary coupling between two modes in presence of the same loss in both modes. Such a simple system is actually a tutorial workhorse in non-Hermitian photonics. For example, adding a local dephasing with the rate $\Gamma>0$ to one of the modes, say, taking 
\begin{eqnarray}
\frac{d}{dt}\vec{\phi}^{(1)}=-i\begin{pmatrix} 
-i(\gamma_1+\Gamma) & v_1 \\
v_1^* & -i\gamma_1
\end{pmatrix}\vec{\phi}^{(1)},
\label{2am0}
\end{eqnarray}
one can get a system with an exceptional point and $\mathcal{PT}$-symmetry (or breaking of it) \cite{PhysRevA.96.053845,Peshko:23}. Our off-diagonal disorder does not induce coupling between amplitudes. It is acting just like local dephasing, i.e. diagonal disorder. For this case also the character of disorder is not relevant.  Correlations of non-conjugated noise variables (i.e., like $\langle \mu_j(t)\mu_j(\tau)\rangle_c$ ) do not appear while deriving Eqs.(\ref{2am}).

The situation for the two-operator averages is quite different.  For the vector of two-operator averages 
\[
\left[\vec\phi^{(2)}\right]^T=[\langle a^{\dagger}_1a_1\rangle,\langle a^{\dagger}_2a_2\rangle,\langle a^{\dagger}_1a_2\rangle,\langle a^{\dagger}_2a_1\rangle],
\]
one has
\begin{eqnarray}
\frac{d}{dt}\vec{\phi}^{(2)}=\begin{pmatrix} 
-2\gamma_1 & 2\gamma_1& -iv_1&iv_1^* \\
2\gamma_1 & -2\gamma_1 & iv_1 & -iv_1^*\\
-iv_1^* & iv_1^* &-2\gamma_1 & 2\kappa_1^* \\
iv_1 & -iv_1 & 2\kappa_1 & -2\gamma_1
\end{pmatrix}\vec{\phi}^{(2)}.
\label{3am}
\end{eqnarray}
The system (\ref{3am}) shows that even for zero average interaction constant (i.e., $v_1=0$) coupling arises between waveguides. They exchange energy and divide it equally. Moreover, for $v_1\neq 0$ the asymptotic result does not depend on $v_1$. The systems indeed act as a grinder equally dividing the energy between the waveguides. Figs.~\ref{fig0-4}(c,d) demonstrates that for the fixed $\gamma_1$ the "grinding" occurs for approximately the same time for quite different values of the  interaction constant $v_1$ and the rate $\kappa_1$. For Fig.~\ref{fig0-4}(c)   all the curves initially the mode $a_1$ is in the coherent state with the amplitude 1, the initial state of the mode $a_2$ is vacuum. 

However, the modal intensity noises (i.e. widths of the modal photon number distributions) are far from being so robust and impervious as intensities are. Respective intensity noise of a $j$-th mode is characterized by the normalized second-order correlations function
\begin{equation}
g^{(2)}_j(t)=\frac{\langle(a_j^{\dagger}(t))^2a_j(t)^2\rangle}{\langle a_j^{\dagger}(t)a_j(t)\rangle^2}.
\label{g2}
\end{equation}
 For two-mode problem it is possible to derive from Eq.(\ref{mast1}) a closed system of equations for the nine-component vector of four-operator averages (\ref{f3a}) .
This system is given in the Appendix A by Eqs.(\ref{4ama}). Results for the correlations functions $g_j(t)^{(2)}$ of both modes are shown in Fig.~\ref{fig0-4}(d) for the  same values of $v_1$, $\kappa_1$ as for Fig.~\ref{fig0-4}(c). Some features remain similar to those of the intensity dynamics. Namely, noise is eventually equalized in both modes, i.e., $g^{(2)}_1(t\rightarrow\infty)=g^{(2)}_2(t\rightarrow\infty)$. However, whereas the initially coherently excited mode always starts with $g^{(2)}_1(0)=1$ typical for the coherent states, the second initially empty mode instantly jumps to $g^{(2)}_2(t\rightarrow 0)\ge 2$ typical for superbunched (super-thermal) states. The value $g^{(2)}_2(t\rightarrow 0)$ depends on the character of off-diagonal random disorder. For our example of real $\kappa_1$ changing from $0$ to $\gamma_1$, the values of  $g^{(2)}_2(t \rightarrow0)$ change  from the thermal value of $2$ to the super-bunched value of $3$. Also, asymptotically the values of $g_j^{(2)}(t)$ tend to the value higher than unity, and in absence of unitary coupling ($v_1=0$) the value of $g^{(2)}_2$ is never lower than its asymptotic value, $g^{(2)}_2(t)\ge g^{(2)}_2(t\rightarrow\infty)$. It is actually a particular manifestation of the "photonic thermalization gap" discussed in Ref. \cite{2015NatPh..11..930K}.

\subsection{Noise jump and "thermalization" gap}

Here we discuss in more details two interesting phenomena connected with small-time and long-time behavior of $g^{(2)}_{1,2}(t)$ visible in Fig.~\ref{fig0-4}(d) and shown in Fig.~\ref{fig0-4}(b).    

As it was already mentioned in the previous Subsection, two-mode system with off-diagonal random disorder displays the "photonic thermalization gap". However, the actual size of this gap depends on the character of noise. We found that actually the asymptotic values $g^{(2)}_{1,2}(t\rightarrow\infty)$ demonstrate very peculiar behaviour in dependence on the rate $k_1$ and the average interaction constant $v_1$. A lower surface in Fig.~\ref{fig0-4}(b) depicts values of $g^{(2)}_{2}$ for the moment corresponding to $\gamma_1 t=10$ for real $\kappa_1$,$v_1$. Firstly, one can see that the values of $g^{(2)}_{2}$ are independent of $v_1$.  Secondly, almost everywhere the value of $g^{(2)}_{2}$ is close to $4/3$, and only for $\kappa_1$ close to $\gamma_1$ the value 
$g^{(2)}_{2}$ tends to $3/2$. A key to this features can be found in stationary solution of the system (\ref{4ama}). One gets from it
that the sum $P=\langle(a_1^{\dagger})^2a_1^2\rangle+\langle(a_2^{\dagger})^2a_2^2\rangle+2\langle a_1^{\dagger}a_2^{\dagger}a_2a_1\rangle$ does not depend on $t$, and  for $t\rightarrow\infty$ and $\kappa_1\neq \gamma_1$
\begin{eqnarray}
\label{as1}
\langle(a_1^{\dagger})^2a_1^2\rangle=\langle(a_2^{\dagger})^2a_2^2\rangle, \\
\nonumber
(\gamma_1^2-\kappa_1^2)(\langle(a_1^{\dagger})^2a_1^2\rangle-2\langle a_1^{\dagger}a_2^{\dagger}a_2a_1\rangle)=0.
\end{eqnarray}
From Eqs.(\ref{as1}) it follows that for $\kappa_1\neq \gamma_1$ the normalized second order correlation function $g^{(2)}_{1,2}(t\rightarrow\infty)=4/3$. 

Only for $\kappa_1= \gamma_1$ one has from Eqs. (\ref{4ama}) the following 
\begin{equation}
\langle(a_1^{\dagger})^2a_1^2\rangle=3\langle a_1^{\dagger}a_2^{\dagger}a_2a_1\rangle
\label{as2}
\end{equation}
and gets $g^{(2)}_{1,2}(t\rightarrow\infty)=3/2$. In practice, for finite intervals  and $\kappa_1$ close to $\gamma_1$, $g^{(2)}_{1,2}$ is going very slowly toward the  value $4/3$, and the behavior shown in Fig.~\ref{fig0-4}(b)  is seen. 

The second feature visible in Fig.~\ref{fig0-4}(d) and shown in Fig.~\ref{fig0-4}(b) is a "noise jump". The upper surface on this plot shows that for the coherent excitation of the first mode,  the photon number noise of initially vacuum second mode suddenly jumps over the thermal values and shows strong dependence on the ratio of $\kappa_1$ and $\gamma_1$. This behaviour can also be inferred from (\ref{4ama}) in the limit $t\rightarrow 0$. For real $v_1$ and $\kappa_1$ one has for the initially vacuum mode
\begin{equation}
g^{(2)}_{2}(t\rightarrow 0)=2+\frac{\kappa_1^2}{\gamma_1^2}.
\label{zer1}
\end{equation}
The parabolic dependence exhibited by Eq.(\ref{zer1}) can be seen in Fig.~\ref{fig0-4}(b). 

\begin{figure*}[ht]
    \begin{minipage}[b]{.3\textwidth}
        \centering
        (a)\\       
        \includegraphics[width=\linewidth]{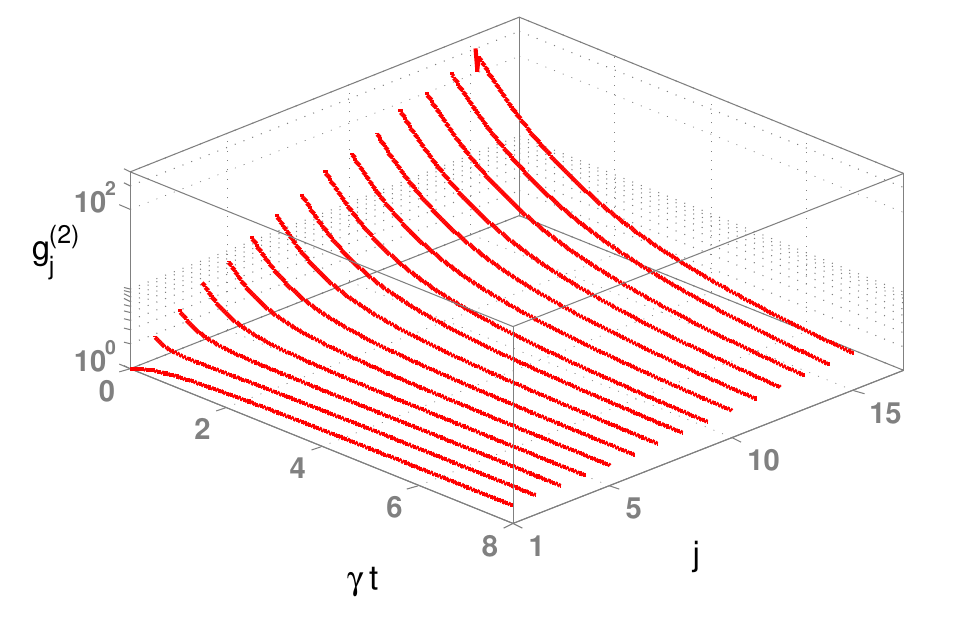} \\        
        (b)\\       
        \includegraphics[width=\linewidth]{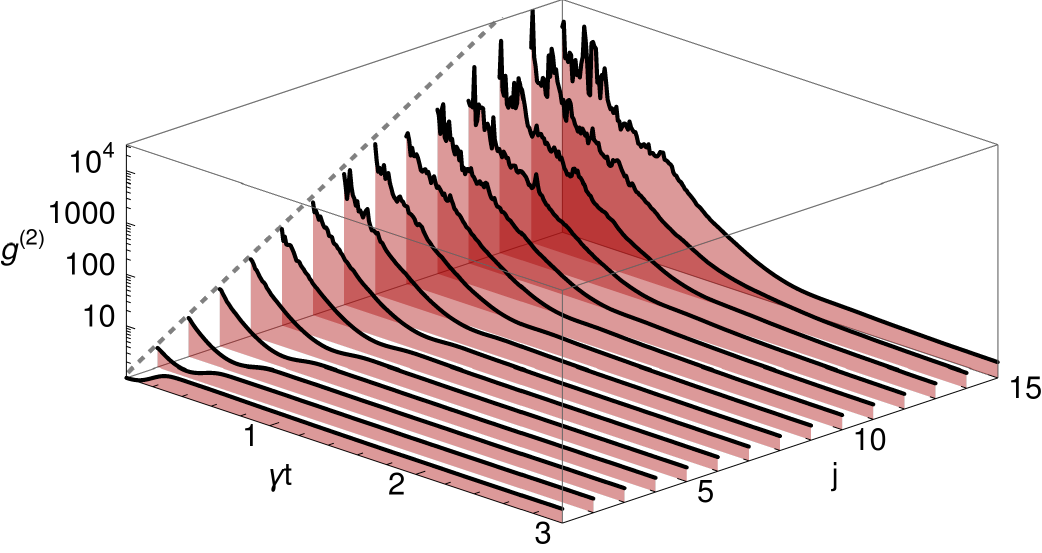}
    \end{minipage}
      \hfill
    \begin{minipage}[b]{.3\textwidth}
        \centering
        (c)\\        
        \includegraphics[width=\linewidth]{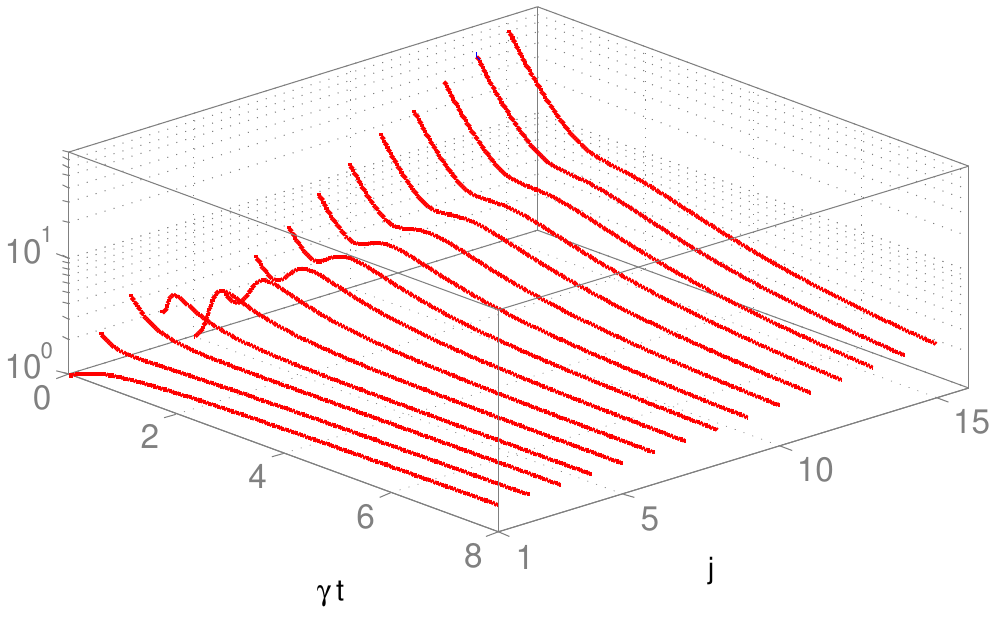}
        (d)\\        
        \includegraphics[width=\linewidth]{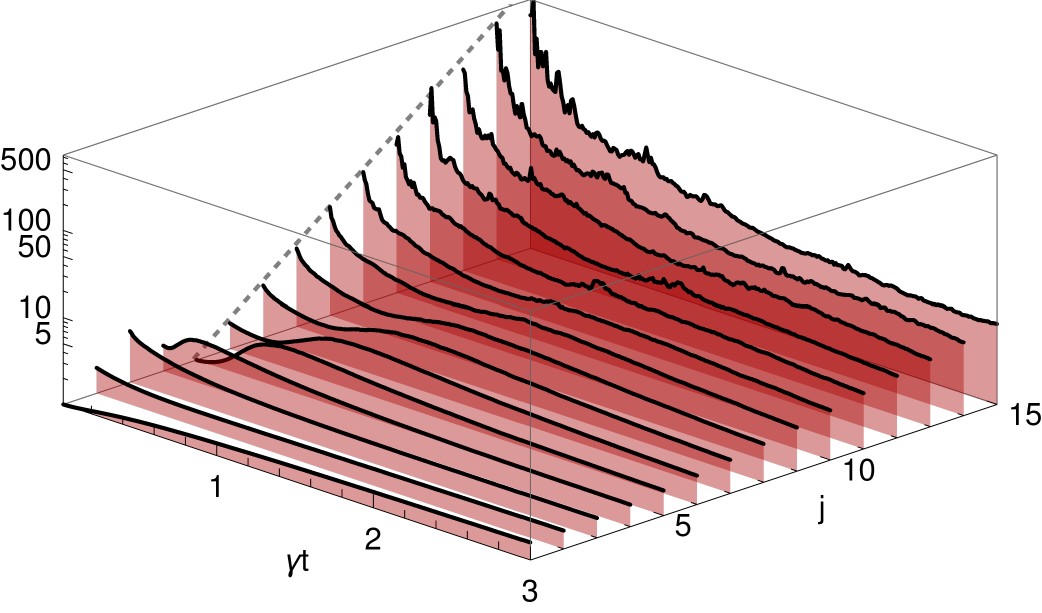}
    \end{minipage}
    \hfill
    \begin{minipage}[b]{.3\textwidth}
        \centering
        (e)\\
        \includegraphics[width=\linewidth]{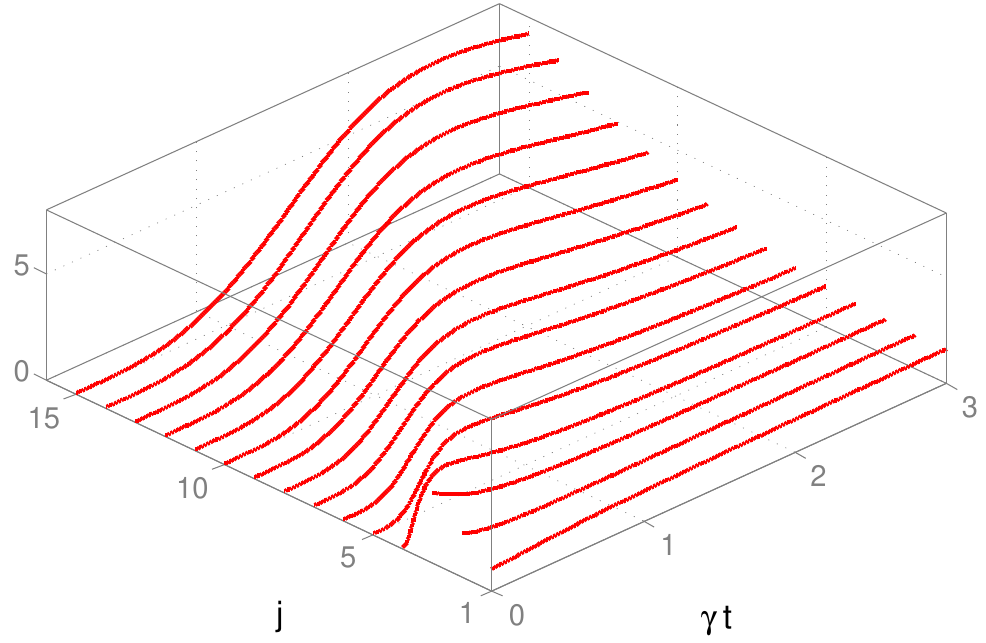}  \\      
        (f)\\        
        \includegraphics[width=\linewidth]{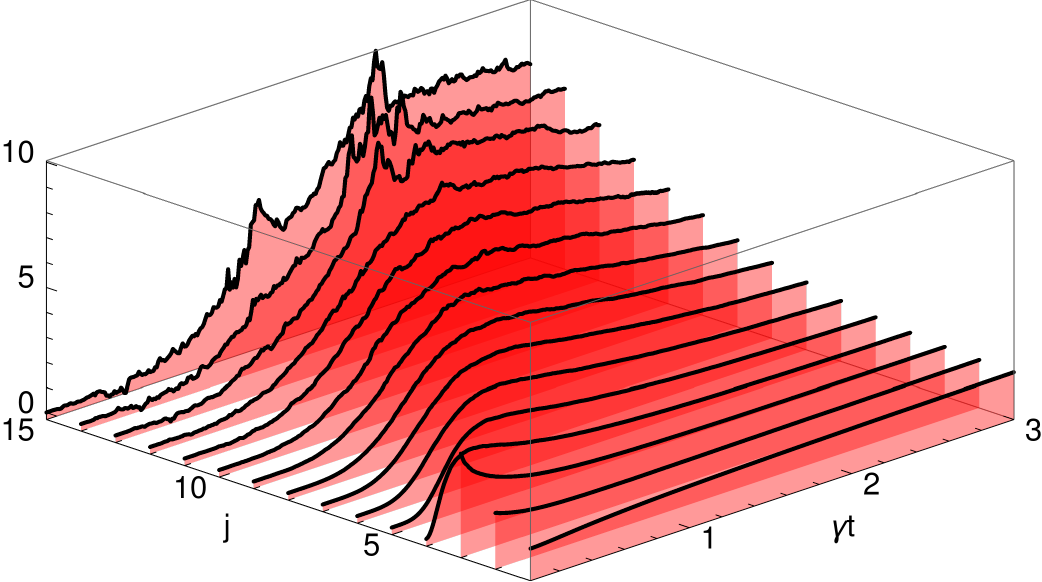}
    \end{minipage}    
\caption{ Normalized second order correlation functions for $j$-th mode  (a,c,e) for the systems described by Eqs.(\ref{n4},\ref{n4k1},\ref{n4k0}) from the master equation approach  and (b,d,f) obtained from the effective Hamiltonian approach, where the number of time sub-intervals is $M=300$ and the number of realization is $K=3*10^5$.  (a,b) The first mode is initially excited in a coherent state with the amplitude $\alpha_1=10$ and $v=5\gamma$, $\nu=0.5$;  (c,d) the first mode is initially excited in a coherent state with the amplitude $\alpha_1=10$ and the fifth mode initially excited with the amplitude $\alpha_5=1$ and $v=0$, $\nu=1$; (e,f) the first mode is initially excited in a coherent state with the amplitude $\alpha_1=10$ and the fifth mode initially excited with the single-photon Fock state; all other unmentioned modes are initially in the vacuum state.  Dashed lines show a curve (b) $2.36^{j-1}$ and (d) $2^{j-5}$ as limit case for $t=0$.
}
\label{fig5-10}
\end{figure*}

\subsection{Effective Hamiltonian approach for two modes}

As it was described  in the subsection IIA, we realized the effective Hamiltonian approach   by generating realizations of the random interaction constants in the time-interval of our interest, finding unitary dynamics of the modal amplitudes for these realizations and than averaging intensities and squared intensities for all the realizations for each chosen time-moment to find $g^{(2)}_j(t)$. To compare the results with ones obtained by the master equation approach, we did it in the following way. We separated the time-interval of our interest on $M\gg 1$ equal sub-intervals and generated in each interval a set of $K$ random values of the interaction constant $v_1$ getting a set of $\{v^{(k)}_{1(m)}\}$ values of interaction constant for $m$-th sub-interval and $k$-th realization. For the $k$-th realization the values of modal coherent amplitudes are propagated as
\begin{equation}
\begin{bmatrix} \alpha_1(t_{m+1}) \\
\alpha_2(t_{m+1})\end{bmatrix}=e^{-i(t_{m+1}-t_m)H_m^{(k)} }\begin{bmatrix} \alpha_1(t_{m}) \\
\alpha_2(t_{m})\end{bmatrix},
\label{mc1}    
\end{equation}
where $t_m$ corresponds to the beginning of the $m$-th interval, and $t_{m+1}$ corresponds to the end of this interval and possible beginning of the next one; the Hamiltonian for the $m$-interval for $k$-th realization is 
\[
H_m^{(k)}=\begin{pmatrix}
0& v_{1(m)}^{(k)} \\
\left(v_{1(m)}^{(k)}\right)^* & 0
\end{pmatrix}.
\]

The result of the illustrative Monte-Carlo simulation is shown in Fig.~\ref{fig0-4}(e,f). For each sub-interval the value of the interaction constant was taken as 
\[
v_{1(m)}^{(k)}=\bar{v}_1+x+i\nu y,  
\]
where the sampled values of independent real random variables $x$ and $y$ were assumed as normally distributed with the zero average  and the variance $v^2$.   We choose parameters as to make time-scales of dynamics close to those obtained by the master-equation approach and shown in Figs.~\ref{fig0-4}(c,d). To that end we assume $\gamma_1=v^2\Delta t$, where $\Delta t$ is the size of the time sub-interval used for simulation. All the curves were obtained for $K=5*10^3$ realizations and the initial states as for Figs.~\ref{fig0-4}(c,d). 

The solid curve in Fig.~\ref{fig0-4}(e,f) corresponds to $\bar{v}_1=0$ and $\nu=0$; the dotted curve corresponds to  $\bar{v}_1=15\gamma_1$ and $\nu=0$; the dashed curve corresponds to $\bar{v}_1=5\gamma_1$ and $\nu=0.5$; the dash-dotted curve corresponds to $\bar{v}_1=0$ and $\nu=1$.

From Fig.~\ref{fig0-4}(e,f) one can see that the effective Hamiltonian approach described by Eq.(\ref{mc1}) corroborates the results obtained by the master equation approach. Asymptotic values of $g^{(2)}_j$ show the same dependence on the nature of the off-diagonal random disorder, and the noise jumps in the initially vacuum mode are also present. The "jump" values of $g^{(2)}_j(t\rightarrow 0)$ obtained by the  effective Hamiltonian  approach also seem to be rather close  to the value given by the master equation approach.

\section{Noise avalanche}

Here we discuss a counter-intuitive phenomenon that can be observed in the multi-mode bosonic chain. It is similar to the bunching "jump" discussed in the previous Section. However, in the multi-mode systems this "jump" rapidly grows with the distance from the initially excited mode and turns into the veritable bunching "avalanche". 

\subsection{Noise avalanche with the master equation}

Here we demonstrate an appearance of the "avalanche" with the relatively simple case of  random off-diagonal disorder with  the white circular noise of the interaction constants. 
Also for simplicity sake, here we consider only the case of zero average interaction constants, $v_j=0$, $\forall j$. For $N+1$ coupled waveguides one gets from the master equation  (\ref{mast1}) the following simple equation for the average number of modal photons
\begin{equation}
\frac{d}{dt}n_j =-2(\gamma_{j-1}+\gamma_j)n_j+2\gamma_{j-1}n_{j-1}+2\gamma_jn_{j+1},
    \label{n1}
\end{equation}
where $n_j(t)=\langle a_j^{\dagger}(t)a_j(t)\rangle$ and $1\leq j\leq N+1$. For chain edges one has to assume $\gamma_{j-1}\equiv 0$ for $j<2$ and $\gamma_{j+1}\equiv 0$ for $j>N$. Eq.(\ref{n1}) was considered in Ref.\cite{PhysRevA.94.012116}. It resembles a common equation for time-continuous classical 1D random walk \cite{MULKEN201137} and it is indeed a "grinder": for any non-zero $\gamma_j$ the stationary state corresponds to the initial population equally distributed among modes, 
\[n_j(t\rightarrow\infty)=\frac{1}{N+1}\sum\limits_{l=1}^{N+1}n_l(0), \forall j.\]

For zero average interaction constants it is also possible to obtain a simple closed set of equation for the four-operator averages of the following kind
\[
n^{(k)}_j(t)=\langle a_j^{\dagger}(t)a_{j+k}^{\dagger}(t)a_{j+k}(t)a_j(t)\rangle,
\]
where $j+k<N+2$. These equations are given in the Appendix \ref{sb}.

Examples of the second order correlation functions for the first mode initially excited in a coherent state are shown in Fig.~\ref{fig5-10}(a). Initial states of all other modes are vacuum. Also, we have taken $\forall j$, $\gamma_j\equiv\gamma$. In Fig.~\ref{fig5-10}(a) one can see that indeed the noise "jump" in behavior of $g_j^{(2)}(t\rightarrow 0)$ is indeed becomes an "avalanche" with increasing $j$. Log-scaled plot of Fig.~\ref{fig5-10}(a) shows  an exponential-like dependence of $g_j^{(2)}(t\rightarrow 0)$ on the distance from the initially excited mode.  

Also, one should note that for$N\gg1$ the asymptotic statistics looks rather close to the thermal one. Indeed, for the circular white noise disorder it is easy to get from Eqs.(\ref{n4},\ref{n4k1},\ref{n4k0}) of the Appendix \ref{sb} that 
\[
g_j^{(2)}(t\rightarrow\infty)\rightarrow \frac{2N}{N+1}
\]
However, for the modes more distant to the initially excited mode, more time is required to approach the limit.

\subsection{Noise avalanche with effective Hamiltonian}

Now let us show how the noise avalanche effect can be reproduced by the effective Hamiltonian approach described by Eqs.(\ref{efs1}). To perform a Monte-Carlo simulation, we are discretizing time  similar to as it  was done for the two-mode system by Eq.(\ref{mc1}).  For each sub-interval the value of the interaction constant between $j$-th and $j+1$-th mode for $m$-th time-interval and $k$-realization  was taken as 
\begin{equation}
v_{j(m)}^{(k)}=\bar{v}_j+x_j+i\nu y_j, 
\label{vN}
\end{equation}
where similarly to the previous Section,  the sampled values of independent real random variables $x_j$ and $y_j$ were assumed as normally distributed with the zero average  and the variance $v^2$. Here we assume $\gamma=v^2\Delta t$, and $\Delta t$ is the size of the time sub-interval used for simulation.

An example of the Monte-Carlo simulation with the real off-diagonal random disorder can be seen in Fig.~\ref{fig5-10}(b). 
All the curves were obtained for $K=5*10^5$ realizations and the initial states as for Fig.~\ref{fig5-10}(a); also $N=15$ modes were taken. Here we have depicted the case when the noise is neither circular nor purely real ($\nu=0.5$) and the average interaction constants are non-zero ($\langle v_j\rangle_c=v_0$, $\forall j$; $v_0=5\gamma$); $K=3*10^5$.  One can see that the effective Hamiltonian approach is indeed capturing noise avalanche effect. The normalized second-order correlation function for $t\rightarrow 0$ grows nearly  exponentially with distance from the initially excited mode.

Curiously, the effective Hamiltonian approach offers a simple analytical estimate for the avalanche growth and differences in dependence on the character of noise. Indeed, for the beginning of the dynamics of the particular $k$-th realization of the interaction constant from the system  (\ref{efs1}) it follows for the coherent state amplitude of $j$-th mode
\begin{equation}
\alpha_{j+1}^{(k)}(j\Delta t)\approx \alpha_1 (-i\Delta t)^j \prod\limits_{l=1}^{j}v^{*(k)}_l,
\label{ajkm}    
\end{equation}
where $\alpha_1$ is the initial coherent state amplitude of the first mode. So, for the normalized second-order correlation function one gets from Eq.(\ref{ajkm})
\begin{equation}
g_{j+1}^{(2)}(j\Delta t)\approx \frac{ \prod\limits_{l=1}^{j}\langle|v_l|^4\rangle_c}{\Bigl(\prod\limits_{l=1}^{j}\langle|v_l|^2\rangle_c\Bigr)^2},
\label{g2No}
\end{equation}
Eq.(\ref{g2No}) captures exponential growth of the initial value of the normalized second-order correlation function with distance from the first mode. An influence of the noise character also seems to be qualitatively captured.  

For the case of all the noise being identically normally distributed, from Eqs. (\ref{vN},\ref{g2No}) one obtains
\begin{equation}
    g_{j+1}^{(2)}(j\Delta t)\approx \left(2+\frac{(1-\nu^2)^2}{(1+\nu^2)^2}\right)^j.
\end{equation}
The dashed line in Fig.~\ref{fig5-10}(b) illustrate this representing the dependence  $2.36^j$. Generally, the noise avalanche grows exponentially with a rate between 2 and 3: for the circular noise ($\nu=1$) one gets  $g_{j+1}^{(2)}(j\Delta t)\propto 2^{j}$, the real noise ($\nu=0$) one has  $g_{j+1}^{(2)}(j\Delta t)\propto 3^{j}$.

It is to be noted that Eq.(\ref{g2No}) actually captures the fact of non-instantaneous transition of excitation from one mode to another (the master equation approach cannot do that). 

It is also to be noted that Eq. (\ref{g2No}) points to the fact that that for large chains (say, for $N>10$) it is rather difficult to capture a  noise avalanche effect in implementations and even in the Monte-Carlo simulations.  Indeed, the same logic that leads to Eq. (\ref{g2No}) shows also that the variance of $g_{j+1}^{(2)}(j\Delta t)$ grows like $(j+1)*x^{2j}$, with $x\in [2,3]$. It means that, for example, if for the real noise one one has a certain variance of $g^{(2)}$ for $K$ realizations in the first waveguide, one would need of about  $K*x^{20}$ noise realizations to have the same variance in the tenth waveguide. One can see a manifestation of this effect in Figs.~\ref{fig5-10}(b,d,f). Results of the Monte-Carlo simulation are visibly noisier with the distance to the first waveguide.

\subsection{Quenching the noise avalanche}

Until now we have considered excitation with a classical coherent state  in the first mode and observed the noise avalanche effect: exponential growth of photon number noise with distance from the first mode. Now let us consider what happens when the other modes are excited. Particularly, we consider how an additional initial excitation of an other mode influences the noise avalanche. 

The result of such a two-mode coherent excitation obtained with the master equation approach one can see in Fig.~\ref{fig5-10}(c). The first mode is excited in a coherent state with the amplitude $\alpha_1=10$, and the fifth mode is excited in a coherent state with the amplitude $\alpha_5=1$; all the other parameters as as for Fig.~\ref{fig5-10}(a).  It can be seen in Fig.~\ref{fig5-10}(c) that the avalanche is delayed, but after the fifth mode it develops in a usual exponential way.

However, when the fifth mode is initially excited in a single-photon state, the resulting noise propagation is drastically different  (see Fig.~\ref{fig5-10}(e)). Despite two orders of magnitude difference in the initial number of photons in the first and fifth modes, there is no avalanche at all. For $j\geq3$ the one has $g^{(2)}_j(t\rightarrow0)=0$.  Eventually, the photon number noise does raise up in other modes after some interaction time. However, the achieved superbunching is much lower than for the coherent initial states. Also, the time-region of antibanching extends with the distance from the mode initially excited in the single-photon state.

Monte-Carlo simulation for a non-classical initial state (i.e., a coherent state in the first mode and a Fock state in the fifth mode) within the effective Hamiltonian framework can be realized in the following way. As a consequence of the linearity,  for every $k$-th realization of the interaction constants $v_{j}^{(k)}(t)$, it is possible to find a solution for the  annihilation operators $a_j(t)$ directly from Heisenberg equation with the Hamiltonian (\ref{ham0}-\ref{hamv}) as
\[a_j(t)=\sum\limits_{l=1}^N S_{jl}^{(k)}(t) a_l(0),\]
where $S_{jl}^{(k)}(t)$ is an evolution matrix for the $k$-th realization. Both the second order correlation function $G_j^{(2)}(t)=\langle a_j^\dagger(t)a_j^\dagger(t)a_j(t)a_j(t)\rangle$ and the number of photons $\mathfrak{n}_j(t)=\langle a_j^\dagger(t)a_j(t)\rangle$ can be expressed as functions of the evolution matrix acting on the initial state. 

The result of such a Monte-Carlo simulation for the 
single-photon state in the fifth mode and the coherent state with the amplitude $\alpha_1=10$ in the first mode one can see in Fig.~\ref{fig5-10}(f). The effect of the avalanche quenching depicted in Fig.~\ref{fig5-10}(e) is obviously captured. 

It is interesting that simple "rule-of-thumb" estimation (\ref{ajkm}) allows one to surmise an effect of quenching. Indeed, zero second-order correlation function for the single-photon state implies zero initial second-order correlation functions for all the initially vacuum modes after the one initially excited in the single-photon state.

\section{Implementations and uses}

As it can be seen from the discussion of the previous Section, just to demonstrate an effect of the noise avalanche one can recourse to the practical approach used in Ref. \cite{2015NatPh..11..930K}. There just sets of single-mode waveguides were taken with different randomly chosen inter-waveguide interaction constants. Experiments were done with these sets and the results were summed up.  Another way to realize the noise avalanche  is to make the interaction constant change in time. To devise randomly fluctuating coupling, one can make use   included such phenomena as photorefractive behaviour for creating a photonic lattice (such as, for example, in Ref.\cite{PhysRevE.66.046602}), or liquid crystal photonic waveguide lattices (such as in Ref.\cite{vitelli2021}).

Another possible way to demonstrate noise avalanche is to implement an analogy between electromagnetic and acoustic wave propagation in waveguides (see, for example, Ref.\cite{Fleury}). The waveguides, which are channels filled with a fluid, e.g. air, support propagation of sound waves in the longitudinal direction, but can be also coupled through the lateral directions if placed close enough \cite{shen2017observation,yin2022characteristics}. Also,  one can imitate fluctuating coupling between neighbour waveguides by modulating fields in one waveguide using the field in the other waveguide, and vice versa {\cite{SIROTA2020108855,10.1063/5.0152144,lasri2023real}.   This can be achieved using feedback-based active acoustic metamaterial waveguides  (see the discussion in the  Appendix \ref{sd}). 

An obvious use of the considered systems with random off-diagonal disorder might be for generation of bunched and super-bunched states in well-defined spatial modes. Such sources are of high demand nowadays, for example, for ghost imaging \cite{PhysRevLett.94.063601,PhysRevLett.89.113601,Shapiro2012ThePO,doi:10.1098/rsta.2016.0233,Simon2017} or   LIDAR ranging based on intensity correlations \cite{Lee2023}, quantum tomography \cite{PhysRevLett.113.070403}. Super-bunched light  can be implemented for enhancing resolution,  contrast and the signal-to-noise ratio for a variety of spatial and temporal imaging schemes \cite{LIU2018824,doi:10.1021/acsphotonics.2c00817}.   

Illustrations of the photon avalanche quenching with single-photon states by Figs.\ref{fig5-10}(e),\ref{fig5-10}(f) point to another curious implementation of the considered scheme. Namely, the scheme could be used to breed sub-Poissonian states. Indeed, in the given illustration at the initial stage of dynamics the states of all the modes apart from the first three ones are sub-Poissonian.

\section*{Conclusions}

Here we considered a tight-binding system of next-neighbor coupled bosonic modes with random off-diagonal disorder. We concentrated our attention on the peculiar phenomenon arising for just one mode initially excited in the coherent state with other modes initially in the vacuum state. This phenomenon is a sudden jump of the photon number noise in the initially empty mode at the very early stages of dynamics. For long  chains of coupled modes this jump turns to the avalanche: the normalized second-order correlation function grows exponentially. We have explained this behavior and derived analytic estimates of the noise growth. We have performed numerical simulation with both the master equation and the effective Hamiltonian approaches demonstrating good similarity between the results. We have envisaged possible realizations of the analyzed systems.  We have also found that single-photon input can suppress the noise avalanche despite the fact that the input coherent state might be of the large number of photons. 

We also envisaged possible applications of the considered system. It might be a useful tool for generating super-bunched states of light for imaging applications. Also, such systems can be used for breeding sub-Poissonian states from a combination of a single-photon and coherent inputs. 

\section*{Funding}
This work is supported by the Swiss National Science Foundation (Grant No. 200021 \_212872). D.M. gratefully acknowledges financial support from the BFFR grants F24MH-001 and F23UZB-064  L. S. was supported by the Israel Science Foundation, grants 2177/23, 2876/23.  

\newpage
\appendix

\section{The system of equations for the two-mode  second-order averages}
For the nine-component vector of four-operator averages  $\vec\phi^{(3)}$ 

    \begin{eqnarray}
    \nonumber
    \left[\vec\phi^{(3)}\right]^T=[\langle(a_1^{\dagger})^2a_1^2\rangle, \langle(a_2^{\dagger})^2a_2^2\rangle,\langle a_1^{\dagger}a_2^{\dagger}a_2a_1\rangle, \\
    \langle(a_1^{\dagger})^2a_2^2\rangle, 
\langle(a_2^{\dagger})^2a_1^2\rangle,\langle(a_1^{\dagger})^2a_1a_2\rangle, \\
\nonumber
\langle a_1^{\dagger}a_2^{\dagger}a_1^2\rangle,\langle(a_2^{\dagger})^2a_1a_2\rangle,\langle a_1^{\dagger}a_2^{\dagger}a_2^2\rangle],
\label{f3a}    
    \end{eqnarray}
one has from Eq.(\ref{mast1}) the following system of equations

    \begin{equation}
\frac{d}{dt} \vec\phi^{(3)}=\begin{pmatrix} \mathcal{A}&\mathcal{B}\\
\mathcal{C}&\mathcal{D}\end{pmatrix}
 \vec\phi^{(3)},
\label{4ama}    
    \end{equation}
where
 \begin{equation}
\mathcal{A}=\begin{pmatrix} 
-4\gamma_1 & 0 &8\gamma_1& -2\kappa_1 & -2\kappa_1^*\\
0 & -4\gamma_1 &8\gamma_1& -2\kappa_1 & -2\kappa_1^* \\
2\gamma_1 & 2\gamma_1 &-8\gamma_1& 2\kappa_1 & 2\kappa_1^*   \\
-2\kappa_1^* & -2\kappa_1^*&8\kappa_1^*& -4\gamma_1 & 0\\
-2\kappa_1 & -2\kappa_1&8\kappa_1& 0& -4\gamma_1 
\end{pmatrix},
\label{A4}    
    \end{equation}

\begin{equation}
\mathcal{B}=\begin{pmatrix} 
-2iv_1 & 2iv_1^* &0&0\\
0&0&-2iv_1^* & 2iv_1\\
iv_1&-iv_1^*&iv_1^* & -iv_1\\
-2iv_1^*&0&0& 2iv_1^*,
\end{pmatrix},
\label{B4}    
    \end{equation}

\begin{equation}
\mathcal{C}=\begin{pmatrix} 
-iv_1^*& 0&2iv_1^*& -iv_1& 0\\
iv_1& 0&-2iv_1&0& iv_1^* \\
0& -iv_1&2iv_1& 0& -iv_1^* \\
0& iv_1^*&-2iv_1^*& iv_1& 0
\end{pmatrix},
\label{C4}    
    \end{equation}

 \begin{equation}
\mathcal{D}=\begin{pmatrix} 
-6\gamma_1&4\kappa_1^{*}&-2\kappa_1^*& 4\gamma_1\\
4\kappa_1&-6\gamma_1&4\gamma_1& -2\kappa_1\\
-2\kappa_1&4\gamma_1&-6\gamma_1& 4\kappa_1\\
4\gamma_1&-2\kappa_1^*&4\kappa_1^*& -6\gamma_1
\end{pmatrix}.
\label{D4}    
    \end{equation}

\begin{figure}
    \centering
    \includegraphics[width=0.8\linewidth]{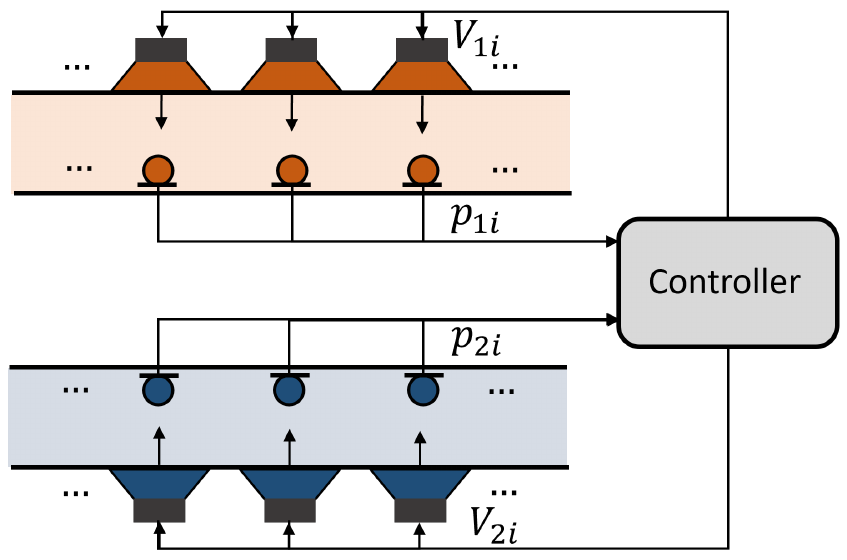}
    \caption{{Active real-time coupling using programmable feedback control - illustration.}}
    \label{fig:Lea}
\end{figure}

\section{The system of equations for the N-mode  second-order averages}
\label{sb}
For zero average interaction constants it is also possible to obtain a simple closed set of equation for the four-operator averages of the following kind
\[
n^{(k)}_j(t)=\langle a_j^{\dagger}(t)a_{j+k}^{\dagger}(t)a_{j+k}(t)a_j(t)\rangle,
\]
where $j+k<N+2$. 

From the master equation (\ref{mast1}) one gets for $k>1$ the following equation
\begin{multline}
\label{n4}
\frac{1}{2}\frac{d}{dt}n^{(k)}_j=-(\gamma_{j-1}+\gamma_j+\gamma_{j+k}+\gamma_{j+k-1})n^{(k)}_j+ 
\gamma_{j-1}n^{(k+1)}_{j-1}+\\+\gamma_{j}n^{(k-1)}_{j+1}+\gamma_{j+k}n^{(k+1)}_{j}+\gamma_{j+k-1}n^{(k-1)}_{j},
\end{multline}
For $k=1$ one has
\begin{align}
    \label{n4k1}
\frac{1}{2}\frac{d}{dt}n^{(1)}_j=-(\gamma_{j-1}+4\gamma_j+\gamma_{j+1})n^{(1)}_j+ 
\gamma_{j}n^{(0)}_{j}+\gamma_{j}n^{(0)}_{j+1}+\gamma_{j-1}n^{(2)}_{j-1}+\gamma_{j+1}n^{(2)}_{j},
\end{align}
and for $k=0$ one obtains
\begin{align}
    \label{n4k0}
\frac{1}{4}\frac{d}{dt}n^{(0)}_j=-(\gamma_{j-1}+\gamma_j)n^{(0)}_j+ 
2\gamma_{j-1}n^{(1)}_{j-1}+2\gamma_{j}n^{(1)}_{j}.
\end{align}

\section{Acoustic realization} 
\label{sd}
Here we discuss a possible way to demonstrate noise avalanche is to implement an analogy between electromagnetic and acoustic wave propagation in waveguides using active acoustic metamaterial waveguides {\cite{SIROTA2020108855,10.1063/5.0152144,lasri2023real}.   We consider coupling not via physical proximity, but via active feedback interaction in real time. The scheme involving two acoustic waveguides is shown in Fig. \ref{fig:Lea}.  The waveguides, which may be placed even very far apart, include an array of electrodynamic loudspeakers (actuators) attached to their cladding and facing inwards, and a corresponding array of microphones (sensors) embedded in the opposite wall. The speakers generate acoustic flow velocities based on sound pressure measurements of the microphones, processed by a reprogrammable electronic controller. Each speaker receives pressure measurements from the own and the other waveguide, and creates the required coupling terms. This real-time feedback operation can be described by
\begin{equation}
    \begin{cases}
        p_{1\_tt}(x,t)=c_1^2p_{1\_xx}(x,t)+\sum_i f_{1\_i}(t)\delta(x-x_i) \\
        p_{2\_tt}(x,t)=c_1^2p_{2\_xx}(x,t)+\sum_i f_{2\_i}(t)\delta(x-x_i)
    \end{cases}
\end{equation}
\begin{equation}
\begin{cases}
    f_{1\_i}(t)=-H_1\left(p_{1\_i}(t),p_{1\_{i\pm 1}}(t),p_{2\_i}(t),p_{2\_{i\pm 1}}(t),...\right) \\
    f_{2\_i}(t)=-H_2\left(p_{2\_i}(t),p_{2\_{i\pm 2}}(t),p_{1\_i}(t),p_{1\_{i\pm 1}}(t),...\right)
\end{cases}
\end{equation}
where $f_1,f_2$ are the control signals (scaled time derivatives of the velocities $V_1,V_2$) to the acoustic actuators, and $H_1,H_2$ are programmed controller gain matrices dependent on the frequency, wavenumber and the coupling constant.}

\bibliography{bibnoise}

\end{document}